\documentclass{iaus}
\usepackage{graphicx}

\title[Magnetic fields in O-type stars] 
{Magnetic fields in O-type stars measured with FORS\,1 at the VLT}

\author[S. Hubrig et al.]   
{S.~Hubrig$^1$,
 M.~Sch\"oller$^2$, R.S.~Schnerr$^3$, I.~Ilyin$^4$, H.F.~Henrichs$^5$,
 R.~Ignace$^6$ \and J.F.~Gonz\'alez$^7$}

\affiliation{$^1$ESO, Santiago, Chile;
email: shubrig@eso.org;
$^2$ESO, Garching, Germany;
$^3$Inst.\ for Solar Physics, Royal Swedish Academy of Sciences, Stockholm, Sweden:
$^4$AIP Potsdam, Germany;
$^5$University of Amsterdam, The Netherlands
$^6$East Tennessee State University, Johnson City, USA;
$^7$Complejo Astronomico El Leoncito, San~Juan, Argentina}

\pubyear{2008}
\volume{259}  
\pagerange{1--2}
\date{?? and in revised form ??}
 \jname{Cosmic Magnetic Fields: from Planets,
to Stars and Galaxies } \editors{K.G. Strassmeier, A.G. Kosovichev
\& J. Beckmann, eds.}
\begin{document}

\maketitle

\begin{abstract}
The presence of magnetic fields in O-type stars has been suspected for a long time. The discovery 
of such fields would explain a wide range of well documented enigmatic phenomena in massive stars, 
in particular cyclical wind variability, H$\alpha$ emission variations, chemical peculiarity, 
narrow X-ray emission lines and non-thermal radio/X-ray emission. Here we present the results of 
our studies of magnetic fields in O-type stars, carried out over the last years.

\end{abstract}

\firstsection 
\section{Introduction}
Direct measurements of the magnetic field strength in massive stars using spectropolarimetry 
to determine the Zeeman splitting
of the spectral lines are difficult, since only a few spectral lines are available for 
these measurements, which are usually strongly broadened 
by rapid rotation.
Before our study (\cite[Hubrig \etal\ 2008]{Hubrig08}),
a magnetic field had only been found in the 
three O-type stars, $\theta^1$\,Ori\,C, HD\,155806, and HD\,191612
(\cite [Donati \etal\ 2002]{donati02};  \cite [Hubrig \etal\ 2007]{hubrig07}; 
\cite [Donati \etal\ 2006]{donati06}).

\section{Observations and analysis}\label{sec:observ}

More than 50 spectropolarimetric observations of 15 O-type stars were obtained from 
2005 to 2008 using the multi-mode instrument FORS\,1 at the 8.2-m Kueyen  telescope. 
GRISMs 600B and 600R were used  with the 0.4$^{\prime\prime}$ slit to obtain $R\approx2000$ 
and $R\approx3000$, respectively. 
Longitudinal magnetic fields were measured in two ways: using only the absorption hydrogen Balmer 
lines or using the entire spectrum including all available absorption lines of hydrogen,
He~I, He~II, C~III, C~IV, N~II, N~III, and O~III.
Most of the targets were observed on three or four different nights to take into account 
the strong dependence of the longitudinal magnetic field on rotational 
aspect. 
Interestingly, a large part of the observed stars exhibited a change of polarity over certain nights.
Four stars of our sample, HD\,36879, HD\,148937, HD\,152408, and HD\,164794, 
showed evidence for the presence of a weak magnetic 
field in the measurements using all spectral absorption lines. The uncertainties in the 
mean longitudinal field determination are obtained from the formal uncertainty in the 
linear regression
of $V/I$ versus the quantity
$-\frac{g_{\rm eff}e}{4\pi{}m_ec^2} \lambda^2 \frac{1}{I} \frac{{\mathrm d}I}{{\mathrm d}\lambda} \left<B_z\right> + V_0/I_0$.

\section{Discussion}
 
This is the first time that magnetic field strengths were determined for such
a large sample of stars, with an accuracy comparable to the errors obtained for the three 
previously known magnetic O-type stars, $\theta^1$\,Ori\,C, HD\,155806, and HD\,191612. For
the magnetic Of?p star HD\,191612, \cite{donati06} measured a magnetic field of $\langle B_\mathrm
{z}\rangle=-220\pm38$\,G, by averaging a total of 52 exposures obtained over 4 different nights.
 This is similar 
to our typical errors of a few tens of G.
The new high-resolution observations ($R\approx30$\,$000$) of the O7~V(n) star HD\,36879 and 
the O8fpe star HD\,191612 with the 
SOFIN echelle spectrograph mounted at the 2.56\,m Nordic Optical Telescope indicate the presence 
of weak magnetic fields of positive polarity in both stars. In spite of a rather 
low signal-to-noise ratio
achieved in these observations (S/N$\approx$200--270), caused by bad weather conditions, 
it was still possible to detect Zeeman features at the positions of He~II, 
C~IV, O~II, and N~III lines.
An example of our observation of HD\,191612 is presented in Fig.\,1.
\begin{figure}
\centering
\includegraphics[height=1.8in,width=2.6in,angle=0]{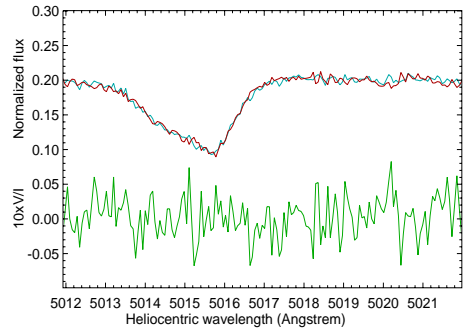}
 \caption{Low signal-to-noise Stokes~$I$ and $V$ spectra of HD\,191612 obtained with the  
echelle spectrograph SOFIN. Due to the rather fast rotation of the star the contribution 
of blends is not easily recognizable in the Stokes~$I$ spectrum. On the 
other hand, the blends become detectable in the Stokes~$V$ spectrum due to the Zeeman signatures 
produced in magnetically sensitive lines. 
 } \label{fig:Ca}
\end{figure}

The four new magnetic O-type stars have different spectral types, luminosity 
classes, and behavior in various observational domains.
The study of the evolutionary state of one of the  Galactic Of?p stars, HD\,191612, indicates 
that it is significantly 
evolved with an $\sim$O8 giant-like classification (\cite [Howarth \etal\ 2007]{how07}).
The youth of the most carefully studied magnetic O-type star $\theta^1$\,Ori\,C and the older age 
of the Of?p star HD\,191612 suggest that the presence of magnetic fields in O-type stars is 
unrelated to their evolutionary state.
We note that it is unclear yet whether more complex, smaller scale fields play a role in 
the atmospheres of hot stars. In the case of a more complex magnetic field topology, 
the longitudinal magnetic field integrated over the visible stellar surface will be smaller 
(or will even cancel) and will not be easily detected with the low-resolution FORS\,1 measurements. 
However, high resolution spectropolarimeters (like ESPaDOnS, Narval, or SOFIN) should 
be able to detect such complex fields.


%



\begin{thebibliography}{}

\bibitem[Donati \etal{} (2002)]{donati02}
Donati, J.-F., Babel, J., Harries, T.~J., et al.\ 2002,
MNRAS, 333, 55

\bibitem[Donati \etal{} (2006)]{donati06}
Donati, J.-F., Howarth, I.~D., Bouret, J.-C., et al.\ 2006,
MNRAS, 365, L6

\bibitem[Howarth \etal{} (2007)]{how07}
Howarth, I.~D., Walborn, N.~R., Lennon, D.~J., et al.\ 2007,
MNRAS, 381, 433

\bibitem[Hubrig \etal{} (2007)]{hubrig07}
Hubrig, S., Yudin, R.~V., Pogodin, M., et al.\ 2007,
AN, 328, 1133

\bibitem[Hubrig \etal{} (2008)]{Hubrig08}
Hubrig, S., Sch{\"o}ller, M., Schnerr, R.~S., et al.\ 2008,
A\&A, 490, 793

\end{thebibliography}
\end{document}